\documentclass[12pt]{article}

\usepackage[utf8]{inputenc}
\usepackage{amssymb,hyperref,amsmath,color}
\usepackage{dsfont}
\usepackage{fullpage}
\usepackage{listings}
\usepackage{graphicx}

\lstnewenvironment{rc}[1][]{\lstset{language=R}}{}

\numberwithin{equation}{section}
\newtheorem{thm}{Theorem}[section]

\allowdisplaybreaks

\DeclareMathOperator*{\argmin}{argmin}

\title{{\sc regMMD}:
\\
An R package for parametric estimation and regression with maximum mean discrepancy
}
\author{Pierre Alquier\footnote{ESSEC Business School},
        Mathieu Gerber\footnote{University of Bristol}}
\date{}

\begin{document}

\maketitle

\begin{abstract}
The Maximum Mean Discrepancy (MMD) is a kernel-based metric widely used for nonparametric tests and estimation. Recently, it has also been studied as an objective function for parametric estimation, as it has been shown to yield   robust estimators~\cite{briol2019statistical,cherief2019finite,alquier2024universal}. We have implemented MMD minimization for parameter inference in a wide range of statistical models, including various regression models, within an \textsf{R} package called {\sc \textbf{regMMD}}. This paper provides an introduction to the {\sc \textbf{regMMD}} package. We describe the available kernels and optimization procedures, as well as the default settings. Detailed applications to simulated and real data are provided.
\end{abstract}

\section{Introduction}
\label{section:intro}

In some models, popular estimators such as the maximum likelihood estimator (MLE) can become very unstable in the presence of outliers. This has motivated research into robust estimation procedures that would not suffer from this issue. Various notions of robustness have been proposed, depending on the type of contamination in the data. Notably, the Huber contamination model~\cite{huber1964robust} considers random outliers while, more recently, stricter notions have been proposed to ensure robustness against adversarial contamination of the data.

The maximum mean discrepancy (MMD) is a kernel-based metric that has received considerable attention in the past 15 years. It allows for the development of tools for nonparametric tests and estimation~\cite{pmlr-v48-chwialkowski16,gretton2007kernel}. We refer the reader to~\cite{muandet2017kernel} for a comprehensive introduction to MMD and its applications. A recent series of papers has suggested that minimum distance estimators (MDEs) based on the MMD are robust to both Huber and adversarial contamination. These estimators were initially proposed for training generative AI~\cite{dziugaite2015training,sutherland2016generative,li2017mmd}, and the study of their statistical properties for parametric estimation, including robustness, was initiated by~\cite{briol2019statistical,cherief2019finite,alquier2024universal}. We also point out that the MMD has been successfully used to define robust estimators that are not MDEs, such as bootstrap methods based on MMD~\cite{dellaporta2022robust} or Approximate Bayesian Computation~\cite{legramanti2022concentration}.

Unfortunately, we are not aware of any software that allows for the computation of MMD-based MDEs. To make these tools available to the statistical community, we developed the {\textsf{R} package called {\sc \textbf{regMMD}}. This package allows for the minimization of the MMD distance between the empirical distribution and a statistical model. Various parametric models can be fitted, including continuous distributions such as Gaussian and gamma, and discrete distributions such as Poisson and binomial. Many regression models are also available, including linear, logistic, and gamma regression. The {\sc \textbf{regMMD}} package is available on the {\sc CRAN} website~\cite{r2013r}:
\url{https://cran.r-project.org/web/packages/regMMD/}

The optimization is based on the strategies proposed by~\cite{briol2019statistical,cherief2019finite,alquier2024universal}. For some models  we have an explicit formula for the gradient of the MMD, in which   case  we   use gradient descent  (see e.g.~\cite{boyd2004convex,nesterov2018lectures}) to perform the optimization. For most models  such a formula does not exist, but we can however   approximate the gradient without bias by Monte Carlo sampling. This allows to use the stochastic gradient algorithm of ~\cite{robbins1951stochastic}, that is one of the most popular estimation methods in machine learning~\cite{bottou-mlss-2004}. We refer to the reader to~\cite{wright2018optimization,duchi2018introductory} and Chapter 5 in~\cite{bach2024learning} for comprehensive introductions to otpimization for statistics and machine learning, including stochastic optimization methods.

The paper is organized as follows. In Section~\ref{section:stat}  we briefly recall the construction of the MMD metric and the MDE estimators based on the MMD. In Section~\ref{section:content}  we detail the content of the package {\sc \textbf{regMMD}}: the available models and kernels, and the optimization procedures used in each case. Finally, in Section~\ref{section:examples}  we provide examples of applications of {\sc \textbf{regMMD}}. Note that these experiments are not meant to be a comprehensive comparison of MMD to other robust estimation procedures. Exhaustive comparisons can be found in~\cite{briol2019statistical,cherief2019finite,alquier2024universal}. The objective is simply to illustrate the use of {\sc \textbf{regMMD}} through pedagogical examples.

\section{Statistical background}
\label{section:stat}

\subsection{Parametric estimation\label{sec:Model_fit}}

Let $X_1,\dots,X_n$ be $\mathcal{X}$-valued random variables identically distributed according to some probability distribution $P^0$, and let $(P_\theta,\theta\in\Theta)$ be a statistical model. Given a metric $d$ on probability distributions, we are looking for an estimator of $\theta_0 \in\argmin_{\theta\in\Theta} d(P_\theta,P^0)$ when such a minimum exists. Letting   $\hat{P}_n=\frac{1}{n}\sum_{i=1}^n \delta_{X_i}$ denote the empirical probability distribution, the  minimum distance estimator (MDE) $\hat{\theta}$ is defined as follows:
$$
\hat{\theta} \in \argmin_{\theta\in\Theta} d(P_\theta,\hat{P}_n ).
$$
The robustness properties of MDEs for well chosen distances was studied as early as in~\cite{wolfowitz1957minimum,parr1980minimum,yatracos2022limitations}. When $ d(P_\theta,\hat{P}_n )$ has no minimum, the definition can be replaced by an $\varepsilon$-approximate minimizer, without consequences on the the properties of the estimator, as shown in to~\cite{briol2019statistical,cherief2019finite}.

Let $\mathcal{H}$ be a Hilbert space, let $\|\cdot\|_{\mathcal{H}}$ and $\left<\cdot,\cdot\right>_{\mathcal{H}}$  denote the associated norms and scalar products, respectively, and let $\varphi:\mathcal{X}\rightarrow\mathcal{H}$. Then, for any probability distribution $P$ on $\mathcal{X}$ such that $\mathbb{E}_{X\sim P}[\|\varphi(X)\|_{\mathcal{H}}]<+\infty $ we can define the mean embedding $\mu(P)=\mathbb{E}_{X\sim P}[\varphi(X)]$. When the   mean embedding $\mu(P)$   is defined for any probability distribution $P$ (e.g.~because the map $\varphi$ is bounded in $\mathcal{H}$), for any probability distributions $P$ and $Q$ we put
$$
\mathbb{D}(P,Q) := \left\| \mu(P) - \mu(Q) \right\|_{\mathcal{H}}.
$$
Letting $k(x,y)=\left<\varphi(x),\varphi(y)\right>_{\mathcal{H}} $, it appears that $\mathbb{D}(P,Q)$ depends on $\varphi$ only through $k$, as we can rewrite:
\begin{equation}
\label{equa:MMD}
\mathbb{D}^2(P,Q)
= \mathbb{E}_{X,X'\sim P} [k(X,X')] -2 \mathbb{E}_{X\sim P,X'\sim Q} [k(X,X')] + \mathbb{E}_{X,X'\sim Q} [k(X,X')].
\end{equation}
When $\mathcal{H}$ is actually a RKHS for the kernel $k$ (see~\cite{muandet2017kernel} for a definition), $\mathbb{D}(P,Q)$ is called the maximum mean discrepancy (MMD) between $P$ and $Q$. A condition on $k$ (universal kernel) ensures that the map $\mu$ is injective, and thus that $\mathbb{D}$ satisfies the axioms of a metric. Examples of universal kernels are known, such as the Gaussian kernel $k(x,y)=\exp(-\|x-y\|^2/\gamma^2)$ or the Laplace kernel $k(x,y)=\exp(-\|x-y\|/\gamma)$, see~\cite{muandet2017kernel} for more examples and references to the proofs.

The properties of MDEs based on MMD were studied in~\cite{briol2019statistical,cherief2019finite}. In particular, when the kernel $k$ is bounded, this estimator enjoys very strong robustness properties. We cite the following very simple result.
\begin{thm}[Special case of Theorem 3.1 in~\cite{cherief2019finite}]
 Assume $X_1,\dots,X_n$ are i.i.d. from $P^0$. Assume the kernel $k$ is bounded by $1$. Then
 $$
 \mathbb{E} \left[ \mathbb{D}\left( P_{\hat{\theta}}, P^0 \right) \right]
 \leq \inf_{\theta\in\Theta} \mathbb{D}\left( P_{\theta}, P^0 \right) + \frac{2}{\sqrt{n}}.
 $$
\end{thm}

Additional non-asymptotic results can be found in~\cite{briol2019statistical,cherief2019finite}. In particular, Theorem 3.1 of~\cite{cherief2019finite} also covers non independent observations (time series). An asymptotic study of $\hat{\theta}$, including conditions for asymptotic normality, can be found in~\cite{briol2019statistical}. All these works provide strong theoretical evidence that $\hat{\theta}$ is very robust to random and adversarial contamination of the data, and this is supported by empirical evidence.

\subsection{Regression\label{sec:reg}}

Let us now consider a regression setting: we observe $(X_1,Y_1),\dots,(X_n,Y_n)$ in $\mathcal{X}\times \mathcal{Y}$ and we want to estimate the conditional distribution $P^0_{Y|X=x}$ of $Y$ given $X=x$ for any $x$. %A direct application of the method in the previous section to the random variables $(X_1,Y_1),\dots, (X_n,Y_n)$ would   lead to the estimation of the joint distribution of the pair $(X,Y)$, which is not the objective of regression models.
To this end we consider a statistical model $(P_\beta,\beta\in\mathcal{B})$ and model the conditional distribution of $Y$ given $X=x$ by $(P_{\beta(x,\theta)},\theta\in\Theta)$ where $\beta(\cdot,\cdot)$ is a specified function $\mathcal{X}\times\Theta \rightarrow \mathcal{B}$. The first estimator proposed by~\cite{alquier2024universal} is:
$$
\hat{\theta}_{\mathrm{reg}}\in\argmin_{\theta\in\Theta} \mathbb{D}\left(
\frac{1}{n}\sum_{i=1}^n \delta_{X_i} \otimes P_{\beta(X_i,\theta)},
\frac{1}{n}\sum_{i=1}^n \delta_{X_i} \otimes \delta_{Y_i}
\right)
$$
where $\mathbb{D}$ is the MMD defined by a product kernel, that is  a kernel of the form $k((x,y),(x',y'))=k_X(x,x')k_Y(y,y')$ (non-product kernels are theoretically possible, but not implemented in the package). Asymptotic and non-asymptotic properties of $\hat{\theta}$ are studied in~\cite{alquier2024universal}. The computation of $\hat{\theta}$ is however slow when the sample size $n$ is large, as it can be shown that  the criterion defining this estimator is the sum of   $n^2$ terms.

By contrast, the following alternative estimator 
$$
\tilde{\theta}_{\mathrm{reg}}\in\argmin_{\theta\in\Theta} \frac{1}{n}\sum_{i=1}^n \mathbb{D}\left(
\delta_{X_i} \otimes P_{\beta(X_i,\theta)},
\delta_{X_i} \otimes \delta_{Y_i}
\right),
$$
has the advantage to be defined through a criterion which is a sum of only $n$ terms. Intuitively, this estimator can be interpreted as a special case of $\hat{\theta}_{\mathrm{reg}}$ where $k_X(x,x')=\mathbf{1}_{ \{x=x'\}}$. An asymptotic study of $\tilde{\theta}_{\mathrm{reg}}$ is provided by~\cite{alquier2024universal}. The theory and the experiments suggest that both estimators are robust, but that $\hat{\theta}_{\mathrm{reg}}$ is more robust than $\tilde{\theta}_{\mathrm{reg}}$. However, for computations reasons, for large sample sizes $n$ ($n>5\,000$, say) only the latter estimator  can be computed in a reasonable amount of time.

\section{Package content and implementation}
\label{section:content}

The package {\sc \textbf{regMMD}} allows to compute the above estimators  in a large number of classical models. We first provide an overview of the two main functions with their default settings, and then give some details on their implementations and   parameters. These functions have many options related to the choice of kernels, the choice of the bandwidth parameters and of the parameters of the optimization algorithms used to compute the estimators. To save space, in this section we only discuss the options that are fundamental from a statistical perspective. We refer the reader to the package documentation for a full description of the available options.

 %We then provide more details on their implementations and    parameters. 
\subsection{Overview of the function  {\tt mmd\_est}}

The function {\tt mmd\_est} performs parametric estimation   as described in Section \ref{sec:Model_fit}. Its required arguments are  the data {\tt x} and the type of model {\tt model} (see Section \ref{sec:model} for the list of available models). Each model implemented in the package has one or two  parameters, namely {\tt par1} and {\tt par2}. If the   model   contains a parameter   that is fixed  (i.e.~not estimated from the data) then its value must be specified by the user. On the other hand, a value for a parameter   that we want to estimate from the data  does not have to be given as an input. If, however,  a value    is provided then   it   is used to initialize  the optimization algorithm that serves at computing the estimator (see below). Otherwise  an initialization by default is used. %When possible, we use gradient descent to compute the estimator of the model parameter, otherwise stochastic gradient is used.

For example, there are three Gaussian univariate models: {\tt Gaussian.loc}, {\tt Gaussian.scale} and {\tt Gaussian}. In each model {\tt par1} is the mean and {\tt par2} is the standard deviation.  In the {\tt Gaussian} model the two parameters {\tt par1} and {\tt par2} are estimated from the data and the MMD estimator of $\theta=$({\tt par1}, {\tt par1}) can be computed as follows:
\begin{verbatim}
R> require("regMMD")
R> estim <- mmd_est(x,model="Gaussian")
\end{verbatim}
By contrast, if we enter
\begin{verbatim}
R> estim <- mmd_est(x,model="Gaussian",par1=0,par2=1)
\end{verbatim}
we simply enforce the optimization algorithm that serves at computing the estimator to use $\theta_0=(0,1)$ as starting value. In the {\tt Gaussian.loc} model only the location parameter (the mean) is estimated. Thus, to compute $\theta=$ {\tt par1} it is necessary to specify the standard deviation {\tt par2}. For instance,
\begin{verbatim}
R> estim <- mmd_est(x,model="Gaussian.loc",par2=1)
\end{verbatim}
will estimate $\theta$ in the model $\mathcal{N}(\theta,1)$. If we provide a value for {\tt par1} then  the optimization algorithm used to compute $\hat{\theta}$ will use $\theta_0={\tt par1}$ as starting value. Finally, In the {\tt Gaussian.scale} model only the scale parameter (standard deviation) is estimated. That is, to estimate  $\theta={\tt par2}$ in e.g.~the $\mathcal{N}(4,\theta^2)$ distribution we can use
\begin{verbatim}
R> estim <- mmd_est(x,model="Gaussian.scale",par1=4)
\end{verbatim}

\subsection{Overview of the function  {\tt mmd\_reg}}

The function {\tt mmd\_reg} is used for regression models (Section \ref{sec:reg})  and requires to specify two arguments, namely the output vector {\tt y} of size $n$ and the $n\times q$ input matrix {\tt X}.  By default, the functions performs linear regression with Gaussian error noise (with unknown variance) and the regression model to be used can be changed through the option {\tt model} of {\tt mmd\_reg} (see Section \ref{sec:model} for the list of available models). In addition, by default, if the input matrix  {\tt X} contains no column whose entries are all equal then an intercept is added to the model, that is, a column of 1's   is added to {\tt X}. This default setting can be disabled by setting the option {\tt intercept} of {\tt mmd\_reg} to {\tt FALSE}.

All regression models implemented in the package have a  parameter   {\tt par1}, and some of them have an additional scalar parameter {\tt par2}. If a model has {\tt par1} as unique   parameter  then the parameter to be estimated is $\theta={\tt par1}$ and  the conditional distribution of $Y$ given $X=x$ is modelled using a model of the form $(P_{\beta(x^\top\theta)},\theta\in\mathbb{R}^k)$, with    $k$ the size of $x$. For instance, for Poisson regression the distribution $P_{\beta(x^\top\theta)}$ is the Poisson distribution with mean $\exp(x^\top\theta)$. By contrast, some  models  have an additional parameter {\tt par2} that also needs to be estimated from the data, so that  $\theta=({\tt par1},{\tt par2})$. For these models  the conditional distribution of $Y$ given $X=x$ is modelled using a model of the form $(P_{\beta(x^\top\gamma,\psi)},\theta=(\gamma,\psi)\in\mathbb{R}^{k+1})$. For instance, for the Gaussian linear regression model with unknown variance   $P_{\beta(x^\top\gamma,\psi)}=\mathcal{N}(x^\top\gamma,\psi^2)$. Finally, some  models  have an additional parameter {\tt par2} whose value needs to be specified by the user. For these models     $\theta= {\tt par1}$ and  the conditional distribution of $Y$ given $X=x$ is modelled using a model of the form $(P_{\beta(x^\top\theta,{\tt par2})},\theta\in\mathbb{R}^{k})$. For instance, for the Gaussian linear regression model with  known variance,  $P_{\beta(x^\top\theta,{\tt par2})}=\mathcal{N}(x^\top\theta,{\tt par2}^2)$. Remark that for all models implemented in the package {\tt par1} is   therefore the vector of regression coefficients. As with the function {\tt mmd\_est}, if a value  for a parameter that needs to be estimated from the data is provided then it   is used to initialize  the optimization algorithm that serves at computing the estimator, otherwise  an initialization by default is used. It is important to note that the number $k$ of regression coefficients of a model is either $q$ (the    number of columns of the input matrix {\tt X}) or $q+1$  if an intercept has been added by {\tt mmd\_reg}. In the latter case, if we want to  provide a value for {\tt par1} then it must be vector of length $q$.

For example, there are two linear regression model with Gaussian noise: {\tt linearGaussian} which assumes that noise variance is unknown and {\tt linearGaussian.loc} which assumes that noise variance is unknown. By default, the former model is used by {\tt mmd\_reg}, and thus linear regression can be simply performed using:
\begin{verbatim}
R> estim <- mmd_reg(y,X)
\end{verbatim}
By default  {\tt mmd\_reg} uses the MMD estimator $\tilde{\theta} _{\mathrm{reg}}$, which we recall is cheaper to compute that the alternative MMD estimator $\hat{\theta}_{\mathrm{reg}}$.
 
\subsection{Kernels and bandwidth parameters\label{sec:ker}}

For parametric models (Section \ref{sec:Model_fit}) the MMD estimator of $\theta$ is computed with a kernel $k(x,x')$ of the form $k(x,x')=K(\|x-x'\|/\gamma)$  for some  bandwidth parameter $\gamma>0$ and some function $K:[0,\infty)\rightarrow [0,\infty)$. The choice of $K$ and $\gamma$ can be specified through the option   {\tt kernel} and \texttt{bdwth} of the function \texttt{mmd\_est}, respectively.  By default, the median heuristic is used to choose $\gamma$ (the median heuristic was used successfully in many applications of MMD, for example~\cite{gretton2012kernel}, see also~\cite{garreau2017large} for a theoretical analysis). The following three options are available for the function  $K$:
\begin{itemize}
 \item \texttt{"Gaussian"}: $ K(u) = \exp(-u^2) $,
 \item \texttt{"Laplace"}: $K(u)=\exp(-u)$,
 \item \texttt{"Gaussian"}: $K(u)=1/(2+u^2)$.
\end{itemize} 

Similarly, for regression models (Section \ref{sec:reg}), the MMD estimators are computed with $k_X(x,x')=K_X(\|x-x'\|/\gamma_X)$ and $k_Y(y,y')=K_Y(\|y-y'\|/\gamma_Y)$  for some bandwidth parameters $\gamma_X\geq 0$ and $\gamma_Y>0$ and   functions $K_X, K_Y:[0,\infty)\rightarrow [0,\infty)$. The choice of $K_X$, $K_Y$, and $\gamma_X$ and $\gamma_Y$ can be specified through the option   {\tt kernel.x}, {\tt kernel.y},  \texttt{bdwth.x}   and \texttt{bdwth.y} of the function \texttt{mmd\_reg}, respectively. The available choices for $K_X$ and $K_Y$ are the same as for $K$, and by default  the median heuristic is used to select $\gamma_Y$. By default,  \texttt{bdwth.x=0} and thus  it is  the  estimator $\tilde{\theta}_{\mathrm{reg}}$ that is used by {\tt mmd\_reg}. The alternative estimator $\hat{\theta}_{\mathrm{reg}}$ can be computed either by providing a positive value for \texttt{bdwth.x} or by setting \texttt{bdwth.x="auto"}, in which case a rescaled version of the  median heuristic is used to choose a positive value for $\gamma_X$.

%
%
%
%  As discussed in Section \ref{sec:reg}, these two estimators rely on product kernels, of the form $k((x,y),(x',y'))=k_X(x,x')k_Y(y,y')$ and in this package,    $k_X(x,x')=K_X(\|x-x'\|/\gamma_X)$ for some bandwidth parameter $\gamma_X\geq 0$ and function $K_X(\cdot)$ (see Section \ref{sec:ker}). Recalling from that  we obtain the estimator $\tilde{\theta}_{\mathrm{reg}}$ when $k_X(x,x')=\mathbf{1}_{ \{x=x'\}}$, it follows that in  {\tt mmd\_reg} choosing $\tilde{\theta}_{\mathrm{reg}}$ amounts to choosing $\gamma_X=0$. The value of this parameter is specified through the option \texttt{bdwth.x} of {\tt mmd\_reg}, and thus the above {\tt R} code is equivalent to
% \begin{verbatim}
% R> estim <- mmd_reg(y,X,bdwth.x=0).
% \end{verbatim}
% From the above discussion it follows that to compute the alternative estimator $\hat{\theta}_{\mathrm{reg}}$ is suffices to provide a positive value for  \texttt{bdwth.x}. Alternatively, using
% \begin{verbatim}
% R> estim <- mmd_reg(y,X,bdwth.x="auto").
% \end{verbatim}
% will compute $\hat{\theta}_{\mathrm{reg}}$ where a data driven approach, the median heuristic, is used to choose  $\gamma_X$. Finally, if we decide to use the model {\tt linearGaussian.loc} in order to set the noise variance equal e.g.~to 1, the estimator $\hat{\theta}_{\mathrm{reg}}$ with $\gamma_X=0.1$ can be computed as follows:
% \begin{verbatim}
% R> estim <- mmd_reg(y,X,bdwth.x=0.1,par2=1,model="linearGaussian.loc).
% \end{verbatim}

\subsection{Optimization methods\label{sec:opt}}

Depending on the model, the  package {\sc \textbf{regMMD}} uses either gradient descent ({\tt GD})  or stochastic gradient descent ({\tt SGD}) to compute the estimators. 

More precisely, for parametric estimation it is proven in Section 5 of~\cite{cherief2019finite} that the gradient of $\mathbb{D}^2(P_\theta,\hat{P}_n)$ with respect to $\theta$ is given by
\begin{equation}
\label{formula:gradient}
  \nabla_\theta \mathbb{D}^2(P_\theta,\hat{P}_n)
 =
 2 \mathbb{E}_{X,X'\sim P_\theta} \left[ \left( k(X,X') - \frac{1}{n}\sum_{i=1}^n k(X_i,X) \right) \nabla_\theta[\log p_\theta(X) ]\right]
\end{equation}
under suitable assumptions on $P_\theta$, including the existence of a density $p_\theta(X)$ and its differentiability. In some models  there is an explicit formula for the expectation in~\eqref{formula:gradient}. This is for instance the case  for the Gaussian mean model, and for such models  a gradient descent algorithm is used to compute the MMD estimator. For models where we cannot compute explicitly the  expectation in \eqref{formula:gradient}  it is possible to compute an unbiased estimate of the gradient by sampling from $P_\theta$. In this scenario  the MMD estimator is computed  using  AdaGrad \cite{Duchi}, and adaptive step-size SGD algorithm. Finally, in very specific models, $\mathbb{D}(P_\theta,\hat{P}_n)$ can be evaluated explicitly in which  case  we can perform exact optimization ({\tt exact}). This is for example  the case when $P_\theta$ is the (discrete) uniform distribution on $\{1,2,\dots,\theta\}$.

In {\tt mmd\_est},  for each model all the available methods for computing the estimator are implemented. The method used by default is chosen according to the ranking: {\tt exact}$>${\tt GD}$>${\tt SGD}. We can enforce another method with the {\tt method} option. For instance,
\begin{verbatim}
R> estim <- mmd_est(x,model="Gaussian")
\end{verbatim}
is equivalent to
\begin{verbatim}
R> estim <- mmd_est(x,model="Gaussian",method="GD")
\end{verbatim}
and we can enforce the use of SGD with
\begin{verbatim}
R> estim <- mmd_est(x,model="Gaussian",method="SGD")
\end{verbatim}

For regression models, formulas similar to the one given in \eqref{formula:gradient} for the gradient of the criteria defining the estimators   $\hat{\theta}_{ \mathrm{reg}}$ and $\tilde{\theta}_{ \mathrm{reg}}$ are provided in Section S1 of the supplement of~\cite{alquier2024universal}. For   the two estimators, this gradient can be computed explicitly  for all  linear regression models when  {\tt kernel.y="Gaussian"}  and for the logistic regression model.

For the  estimator $\tilde{\theta}_{\mathrm{reg}}$, and as for parametric estimation, gradient descent is used when the gradient of the objective function can be computed explicitly,   and otherwise the optimization is performed using Adagrad.  In {\tt mmd\_reg} gradient descent is implemented using backtracking line search to select the step-size to use at each iteration, and a stopping rule is implemented to stop the optimization earlier when possible.

The computation  of $\hat{\theta}_{\mathrm{reg}}$ is more delicate. Indeed, the objective function defining this estimator is the sum of $n^2$ terms (see Section \ref{sec:reg}), implying that  minimizing this function using     gradient descent or SGD leads to algorithms for computing the estimator that require $\mathcal{O}(n^2)$ operations per iteration. In the package, to reduce the cost per iteration we implement the strategy proposed in Section S1 of the supplement of~\cite{alquier2024universal}. Importantly, with this strategy the optimization is performed using an unbiased estimate of the gradient of the objective function, even when the gradient  of the $n^2$ terms of objective function can be computed explicitly. It is  however possible to use the explicit formula for these gradients   to reduce the variance of the noisy gradient, which we do   when possible. As for the computation of $\tilde{\theta}_{\mathrm{reg}}$ a stopping rule is implemented to stop the optimization earlier when possible. With this package it is feasible to compute $\hat{\theta}_{\mathrm{reg}}$ in a reasonable amount of time for dataset containing up to a few thousands data points (up-to $n\approx 5\,000$ observations, say).

Finally, we stress that, from a computational point of view,    MMD estimation in regression models  is a much more challenging task than in parametric models for at least two reasons. Firstly, while in the latter task the dimension of the parameter of interest $\theta$ is at most two for the models implemented in this package (see below), in regression the dimension of $\theta$ can be much larger, depending on the number of explanatory variables. Secondly, the objective functions to minimize for regression models are ``more non-linear''. When estimating a regression model it is therefore a good practice to verify that the optimization of the objective function has converged. This can be done by inspecting the sequence of $\theta$ values computed by the optimization algorithm, accessible from the  object \texttt{trace} of \texttt{mmd\_reg}.

\subsection{Available models\label{sec:model}}

\textbf{List of univariate models in {\tt mmd\_est}:}
\begin{itemize}
 \item Gaussian $\mathcal{N}(m,\sigma^2)$: {\tt Gaussian} (estimation of $m$ and $\sigma$), {\tt Gaussian.loc} (estimation of $m$) and {\tt Gaussian.scale} (estimation of $\sigma$),
 \item Cauchy: {\tt Cauchy} (estimation of the location parameter),
 \item Pareto: {\tt Pareto} (estimation of the exponent),
 \item exponential $\mathcal{E}(\lambda)$: {\tt exponential},
 \item gamma $Gamma(a,b)$: {\tt gamma} (estimation of $a$ and $b$), {\tt gamma.shape} (estimation of $a$) and {\tt gamma.rate} (estimation of $b$),
 \item continuous uniform: {\tt continuous.uniform.loc} (estimation of $m$ in $\mathcal{U}[m-\frac{L}{2},m+\frac{L}{2}]$, where $L$ is fixed by the user), {\tt continuous.uniform.upper} (estimation of $b$ in $\mathcal{U}[a,b]$) and {\tt continuous.uniform.lower.upper} (estimation of both $a$ and $b$ in $\mathcal{U}[a,b]$),
 \item Dirac $\delta_a$: {\tt Dirac} (estimation of $a$; while this might sound uninteresting, this can be used to define a ``model-free'' robust location parameter),
 \item discrete uniform $\mathcal{U}(\{1,2,\dots,N\})$: {\tt discrete.uniform} (estimation of $N$),
 \item binomial $Bin(N,p)$: {\tt binomial} (estimation of $N$ and $p$), {\tt binomial.size} (estimation of $N$) and {\tt binomial.prob} (estimation of $p$),
 \item geometric $\mathcal{G}(p)$: {\tt geometric},
 \item Poisson $\mathcal{P}(\lambda)$: {\tt Poisson}.
\end{itemize}

\textbf{List of multivariate models in {\tt mmd\_est}:}
\begin{itemize}
 \item multivariate Gaussian $\mathcal{N}(\mu,U U^T)$: {\tt multidim.Gaussian} (estimation of $\mu$ and $U$), {\tt multidim.Gaussian.loc} (estimation of $\mu$ while $U=\sigma I$ for a fixed $\sigma>0$) and {\tt multidim.Gaussian.scale} (estimation of $U$ while $\mu$ is fixed),
 \item Dirac mass $\delta_{a}$: {\tt multidim.Dirac}.
\end{itemize}

\textbf{List of regression models in {\tt mmd\_reg}:}
\begin{itemize}
 \item linear regression models with Gaussian noise: {\tt linearGaussian} (unknown noise variance) and {\tt linearGaussian.loc} (known noise variance),
 \item exponential regression: {\tt exponential},
 \item gamma regression: {\tt gamma}, or {\tt gamma.loc} when the precision parameter is known,
 \item beta regression: {\tt beta}, or {\tt beta.loc} when the precision parameter is known,
 \item logistic regression: {\tt logistic},
 \item Poisson regression: {\tt poisson}.
\end{itemize}

\subsection{Output}

The output of both {\tt mmd\_reg} and {\tt mmd\_est} is a list that contains error messages (if any), the estimated parameters (if no error prevented their computations) and various information on the model and the estimators. The estimated values can be recovered respectively by {\tt estim\$estimators} (with {\tt mmd\_est}) and {\tt estim\$coefficients}  (with {\tt mmd\_reg}). A {\tt summary} function is available in both cases:
\begin{verbatim}
R> estim <- mmd_reg(y, X)
R> summary(estim)
\end{verbatim}
Examples of such summaries are provided and discussed in the next section.

\section{Detailed examples}
\label{section:examples}

\subsection{Toy example: robust estimation in the univariate Gaussian model}

We start with a very simple illustration on synthetic data. We choose one of the simplest model, namely, estimation of the mean of a univariate Gaussian random variable. The statistical model is $\mathcal{N}(\theta,1)$, which is the \texttt{Gaussian.loc} model in the package. When the sample is stored in a vector \texttt{x}, the code for the estimation with default settings is:
\begin{verbatim}
R> require("regMMD")
R> estim <- mmd_est(x,model="Gaussian.loc",par2=1)
R> summary(estim)
======================== Summary ========================
Model:               Gaussian.loc
---------------------------------------------------------
Algorithm:           GD
Kernel:              Gaussian
Bandwidth:           0.976611092935025
---------------------------------------------------------
Parameters:                   
 
par1: mean -- initialized at 0.0418
      estimated value: 0.0221 
 
par2: standard deviation -- fixed by user: 1
========================================================= 
\end{verbatim}
Note the default settings for the kernel (Gaussian), the bandwidth (computed by median heuristic) and the initialization of the gradient descent on $\theta$ (taken as the median of the data). The user can fix this initialization instead (observe this has no effect on the convergence in this case):
\begin{verbatim}
R> estim <- mmd_est(X,model="Gaussian.loc",par1=2,par2=1)
R> summary(estim)
======================== Summary ========================
Model:               Gaussian.loc
---------------------------------------------------------
Algorithm:           GD
Kernel:              Gaussian
Bandwidth:           0.976611092935025
---------------------------------------------------------
Parameters:                   
 
par1: mean -- initialized at 2
      estimated value: 0.0221 
 
par2: standard deviation -- fixed by user: 1
=========================================================
\end{verbatim}
The user can also impose a different bandwidth and kernel, which will result in a different estimator.
\begin{verbatim}
R> estim <- mmd_est(X,model="Gaussian.loc",par2=1,bdwth=0.6,kernel="Laplace")
R> summary(estim)
======================== Summary ========================
Model:               Gaussian.loc
---------------------------------------------------------
Algorithm:           GD
Kernel:              Laplace
Bandwidth:           0.6
---------------------------------------------------------
Parameters:                   
 
par1: mean -- initialized at 0.0418
      estimated value: 0.0243 
 
par2: standard deviation -- fixed by user: 1
========================================================= 
\end{verbatim}
We end up the discussion on the Gaussian mean example by a toy experiment: we replicate $N=200$ times the simulation of $n=100$ i.i.d.~random variables $\mathcal{N}(-2,1)$ and compare the maximum likelihood estimator (MLE), equal to  the empirical mean, the median, and   the MMD estimator  with a Gaussian and with a Laplace kernel, using in both cases the median heuristic to choose the bandwidth parameter. In a second time, we repeat the same experiment with contamination: two of the $X_i$'s are sampled from a standard Cauchy distribution instead. The mean absolute error (MAE) of each estimator over all experiments are reported in Figure~\ref{table1}. The results are as expected: under no contamination, the MLE is known to be efficient and is therefore the best estimator. On the other hand, the MLE (i.e.~the empirical mean) is known to be very sensitive to outliers. In contrast,  the    MMD estimators and the median are robust estimators of $\theta$. We refer the reader to~\cite{briol2019statistical,cherief2019finite} for more discussions and experiments on the robustness   MMD estimators.
 
 Note that while the median is a natural robust alternative to the MLE in the Gaussian mean model, such an alternative is not always available. Consider for example the estimation of the standard deviation of a Gaussian, with known zero mean: $\mathcal{N}(0,\theta^2)$. We cannot use (directly) a median in this case, while the MMD estimators are available:
 \begin{verbatim}
R> estim <- mmd_est(x,model="Gaussian.scale",par1=0)
R> summary(estim)

======================== Summary ========================
Model:               Gaussian.scale
---------------------------------------------------------
Algorithm:           SGD
Kernel:              Gaussian
Bandwidth:           1.00146068635693
---------------------------------------------------------
Parameters:                   
 
par1: mean -- fixed by user: 0
 
par2: standard deviation -- initialized at 0.9173
      estimated value: 1.0808 
========================================================= 
\end{verbatim}
We repeat similar experiments as above in this case. We let $N$ and $n$ be as above and sample  the uncontaminated   observations from the $\mathcal{N}(0,1)$ distributions. We them compare the MLE to the MMD with Gaussian and Laplace kernel both in the uncontaminated case  and under Cauchy contamination for two observations. The results are reported in Figure~\ref{table2}. The conclusions are completely similar to the ones obtained in the Gaussian location experiments.

We finally provide an example of estimation of both parameters in Gaussian model.
 \begin{verbatim}
R> estim <- mmd_est(x,model="Gaussian")
R> summary(estim)

======================== Summary ========================
Model:               Gaussian
---------------------------------------------------------
Algorithm:           SGD
Kernel:              Gaussian
Bandwidth:           1.00146068635693
---------------------------------------------------------
Parameters:                   
 
par1: mean -- initialized at 0.1426
      estimated value: 0.1101 
 
par2: standard deviation -- initialized at 0.9288
      estimated value: 1.0768 
========================================================= 
\end{verbatim}

 \begin{figure}
 \begin{center}
  \begin{tabular}{|c||c|c|c|c|}
   \hline
   & MLE & MMD (Gaussian kernel) & MMD (Laplace kernel) & median \\
   \hline
   no contamination &  \textcolor{blue}{{\bf 0.0816}} & {\bf 0.0912} & {\bf 0.0895} & \textcolor{red}{{\bf 0.110}} \\
                    & \textcolor{blue}{(0.062)} & (0.072) & (0.078) & \textcolor{red}{(0.079)} \\
   \hline 
   contamination & \textcolor{red}{{\bf 0.1175}} & {\bf 0.0885} & \textcolor{blue}{{\bf 0.0813}} & {\bf 0.0894} \\
       by Cauchy           & \textcolor{red}{(0.171)} & (0.068) & \textcolor{blue}{(0.067)} & (0.075)   \\        
   \hline
  \end{tabular}
  \end{center}
   \caption{Mean Absolute Error ({\bf MAE}) for various estimators in the (univariate) Gaussian mean experiments (between parenthesis: the standard deviation of the absolute error over all experiments). The best MAE performance is highlighted in \textcolor{blue}{blue} and the worst one in \textcolor{red}{red}.}
 \label{table1}
 \end{figure}
 \begin{figure}
 \begin{center}
  \begin{tabular}{|c||c|c|c|}
   \hline
   & MLE & MMD (Gaussian kernel) & MMD (Laplace kernel)  \\
   \hline
   no contamination & \textcolor{blue}{{\bf 0.0533}} & \textcolor{red}{{\bf 0.0676}} &  {\bf 0.0659 } \\
                    & \textcolor{blue}{(0.042)} & \textcolor{red}{(0.051)} & (0.051)  \\
   \hline 
   contamination & \textcolor{red}{{\bf 0.3926}} & {\bf 0.0742} & \textcolor{blue}{{\bf 0.0733}} \\
       by Cauchy           & \textcolor{red}{(1.129)} & (0.056) & \textcolor{blue}{(0.055)} \\
   \hline
  \end{tabular}
  \end{center}
   \caption{Mean Absolute Error ({\bf MAE}) for various estimators in the (univariate) Gaussian scale experiments (between parenthesis: the standard deviation of the absolute error over all experiments). The best MAE performance is highlighted in \textcolor{blue}{blue} and the worst one in \textcolor{red}{red}.}
 \label{table2}
 \end{figure}

\subsection{Robust linear regression\label{sub:Lin_Reg}}

\subsubsection{Dataset and model}

To illustrate the use of the   {\sc \textbf{regMMD}}  package to perform robust linear regression we use the \textsf{R} built-in dataset \texttt{airquality}. The dataset contains daily measurements of four variables related to air quality in New York, namely the ozone concentration (variable \texttt{Ozone}), the temperature (variable \texttt{Temp}), the wind speed (variable \texttt{Wind}) and the solar radiation (variable \texttt{Solar.R}). Observations are reported for the period ranging from the 1st of May 1973 to the 30th of September 1973, resulting in a total of  153 observations.  The dataset contains 42 observations with missing values, which we remove from the sample.

\begin{figure}[!t]
\centering
\includegraphics[scale=0.4]{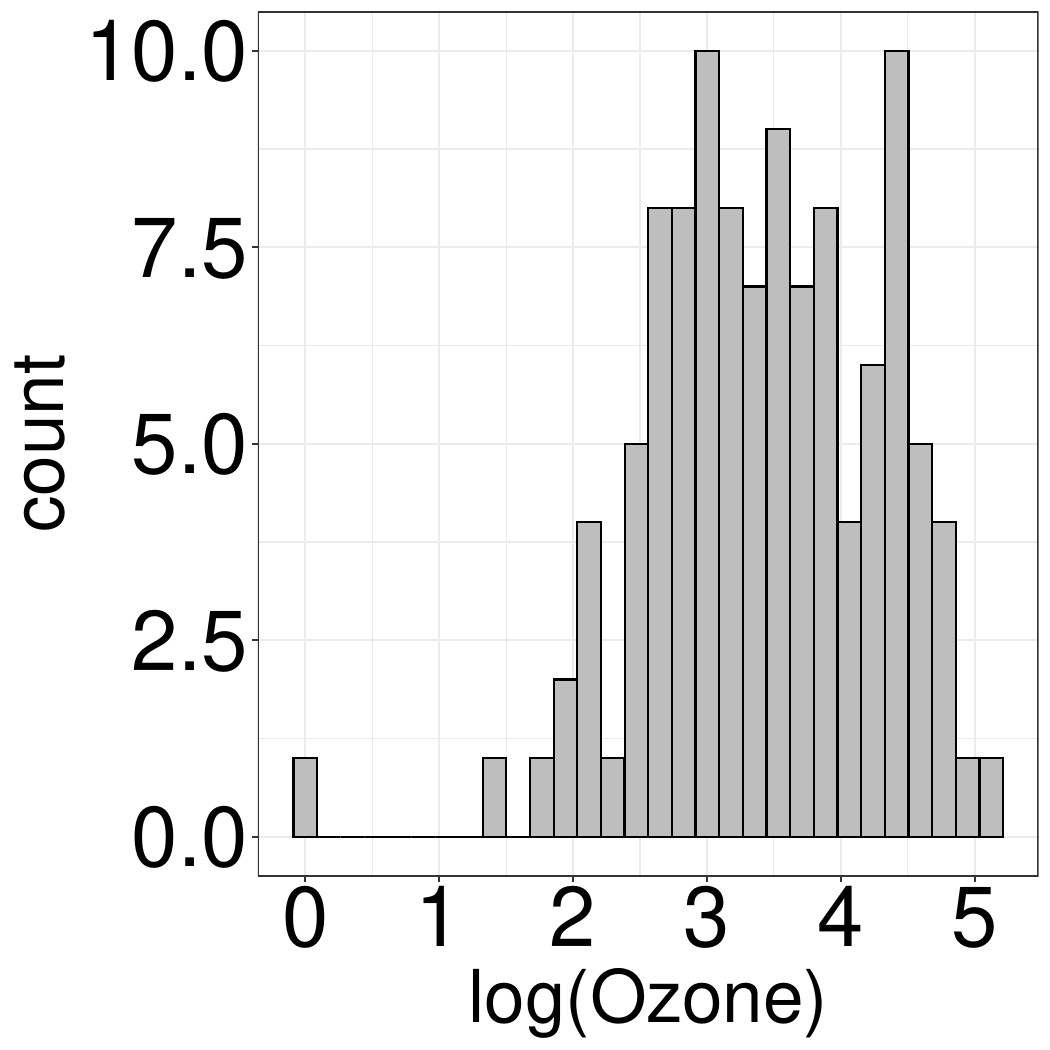}
\hspace{0.2cm}\includegraphics[scale=0.4]{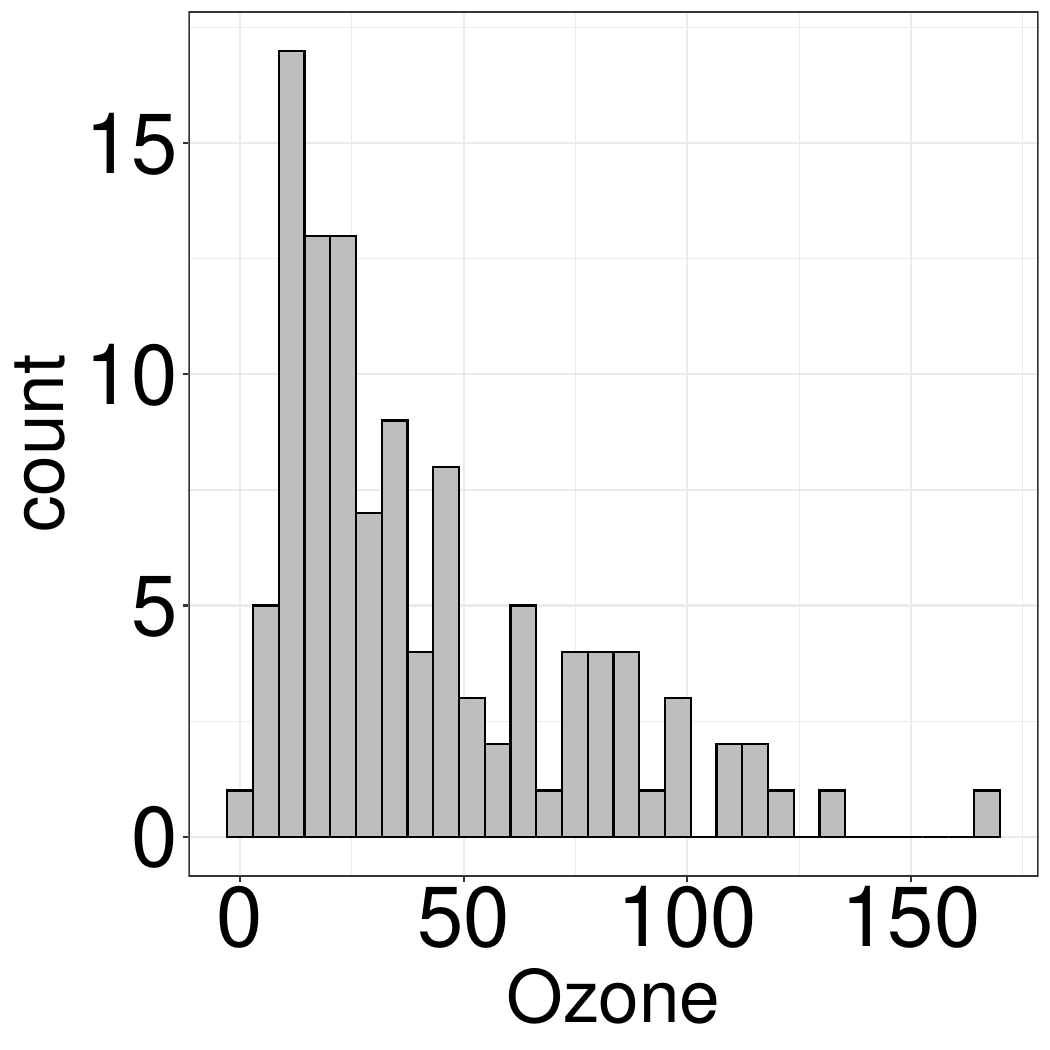}
\caption{\label{figs:ozone1} Distribution of the observed values of $\log(\text{\texttt{Ozone}})$ (left) and of the observed values of \texttt{Ozone} (right).}
\end{figure}

Our aim is to study the link between the level of ozone concentration and the other three variables present in the dataset using the following linear regression model:
\begin{equation*}
\begin{split}
\log(\text{\texttt{Ozone}}) =\alpha&+\beta_1\text{\texttt{Temp}}+\beta_2(\text{\texttt{Temp}})^2+\beta_3 \text{\texttt{Wind}}+\beta_4 (\text{\texttt{Wind}})^2\\
&+\beta_5 \text{\texttt{Solar.R}}+\beta_6 (\text{\texttt{Solar.R}})^2+\epsilon 
\end{split}
\end{equation*}
where $\epsilon\sim\mathcal{N}(0,\sigma^2)$. The noise variance $\sigma^2$ is assumed to be unknown so that the model contains 8 parameters that need to be estimated from the data.

The left plot in Figure \ref{figs:ozone1} shows the distribution of the observed values for the response variable $\log(\text{\texttt{Ozone}})$. From this plot we observe that there is one isolated observation (i.e.~outlier), for which the observed value of $\log(\text{\texttt{Ozone}})$ is  0.

\subsubsection{Data preparation}

We start by loading   the  {\sc \textbf{regMMD}}  package and the dataset:
\begin{verbatim}
R> require("regMMD")
R> air.data <- na.omit(airquality)
\end{verbatim}
and we prepare the vector {\tt y} containing the observations for the response variable as well as the design  matrix {\tt X}:
\begin{verbatim}
R> y <- log(air.data[,1])
R> X <- as.matrix(air.data[,-c(1,5,6)])
R> X <- cbind(poly(air.data[,2], degree=2),poly(air.data[,3], degree=2),
            poly(air.data[,4], degree=2))
\end{verbatim}

\subsubsection{OLS estimation}

We first estimate the model with the ordinary least squares (OLS) approach, using both   the full dataset and the one obtained by removing the outlier:
\begin{verbatim}
R> ols.full <- lm(y~X)
R> ii <- which(y<1)
R> ols <- lm(y[-ii]~X[-ii,])
R> print(cbind(ols.full$coefficients,ols$coefficients))
                 [,1]       [,2]
(Intercept)  3.4159273  3.4361465
X1           2.6074060  2.2885974
X2          -1.2034322 -0.8487403
X1          -2.2633108 -2.5209875
X2           1.2894833  1.1009511
X1           4.2120018  3.9218225
X2           0.4501812  0.8936152

\end{verbatim}

As expected, the OLS estimates are sensitive to the presence of outliers in the data. In particular, we observe that the unique outlier in the data has a non-negligible impact on the estimated regression coefficient of the variables $(\text{\texttt{Temp}})^2 $ and $(\text{\texttt{Solar.R}})^2$.

\subsubsection{MMD estimation with the default settings\label{sub:MMD_default}}
 
Using the default settings of the \texttt{mmd\_reg} function, the computationally cheap  estimator $\tilde{\theta}_{\mathrm{reg}}$ with a Gaussian kernel $k_Y(y,y')$ is used, and the model is estimated as follows:
\begin{verbatim}
R> mmd.tilde <- mmd_reg(y,X)
R> summary(mmd.tilde)
======================== Summary ========================
Model:               linearGaussian
Estimator:           theta tilde (bdwth.x=0)
---------------------------------------------------------
  Coefficients           Estimate
---------------------------------------------------------
  (Intercept)               3.427
  X1                       2.3448
  X2                      -0.8132
  X3                       -2.329
  X4                       1.0565
  X5                       4.1788
  X6                       0.8369
---------------------------------------------------------
  Std. dev. of Gaussian noise : 0.4484 (estimated)
---------------------------------------------------------
  Kernel for y: Gaussian with bandwidth 0.5821
========================================================= 
\end{verbatim}

As expected from the robustness properties of MMD based estimators, we observe that the estimated values of the regression coefficients are similar to those obtained by OLS on the dataset without the outlier. The \texttt{summary} command also  returns the value of the   bandwidth parameter $\gamma_Y$ obtained with the median heuristic and used to compute the estimator. 

To estimate the model using the alternative estimator $\hat{\theta}_{\mathrm{reg}}$ we need to choose a non-zero value for  \texttt{bdwth.x}. Using the default setting, this estimator is computed as follows:
\begin{verbatim}
R> mmd.hat <- mmd_reg(y,X,bdwth.x="auto")
R> summary(mmd.hat)
======================== Summary ========================
Model:               linearGaussian
Estimator:           theta hat  (bdwth.x>0)
---------------------------------------------------------
  Coefficients           Estimate
---------------------------------------------------------
  (Intercept)              3.4269
  X1                       2.3444
  X2                      -0.8131
  X3                      -2.3297
  X4                        1.056
  X5                       4.1786
  X6                       0.8374
---------------------------------------------------------
  Std. dev. of Gaussian noise : 0.4482 (estimated)
---------------------------------------------------------
  Kernel for y: Gaussian with bandwidth 0.5821
  Kernel for x: Laplace with bandwidth 0.0215
========================================================= 
\end{verbatim}

We remark that the  value for  \texttt{bdwth.x} selected by the default setting is very small, and thus the two MMD estimators provide very similar estimates of the model parameters.

\subsubsection{ Tuning the fit in MMD estimation}

Above, the estimator $\hat{\theta}_{\mathrm{reg}}$ was computed using  a Laplace kernel  for  $k_X(x,x')$ and  a Gaussian kernel $k_Y(y,y')$, and using a data driven approach to select their bandwidth parameters $\gamma_X$ and $\gamma_Y$.  If, for instance, we want to use  a Cauchy kernel  for $k_X(x,x')$  with bandwidth parameter $\gamma_X=0.2$  and a Laplace kernel for $k_Y(y,y')$  with bandwidth parameter $\gamma_Y=0.5$, we can proceed as follows:
\begin{verbatim}
R> mmd.hat <- mmd_reg(y,X,bdwth.x=0.1,bdwth.y=0.5,kernel.x="Cauchy",
 					                    kernel.y="Laplace")
R> summary(mmd.hat)
======================== Summary ========================
Model:               linearGaussian
Estimator:           theta hat  (bdwth.x>0)
---------------------------------------------------------
  Coefficients           Estimate
---------------------------------------------------------
  (Intercept)              3.4189
  X1                       2.3633
  X2                      -0.7652
  X3                      -2.3423
  X4                       1.0684
  X5                       4.2731
  X6                       0.8787
---------------------------------------------------------
  Std. dev. of Gaussian noise : 0.4492 (estimated)
---------------------------------------------------------
  Kernel for y: Laplace with bandwidth 0.5
  Kernel for x: Cauchy with bandwidth 0.1
========================================================= 
\end{verbatim}

We recall that different choices for the kernels $k_Y(y,y')$ and $k_X(x,x')$ lead to different MMD estimators, which explains why the estimates obtained here are different from those obtained in Section \ref{sub:MMD_default}.

\subsection{Robust Poisson regression}

\subsubsection{Dataset and model}

As a last example we consider the same dataset and task as in Section \ref{sub:Lin_Reg}, but now assume the following Poisson regression model:
\begin{equation*}
\begin{split}
\text{\texttt{Ozone}} \sim\mathrm{Poisson}\bigg(\exp\Big(\alpha&+ \beta_1\text{\texttt{Temp}}+\beta_2(\text{\texttt{Temp}})^2+\beta_3 \text{\texttt{Wind}}+\beta_4 (\text{\texttt{Wind}})^2\\
&+\beta_5 \text{\texttt{Solar.R}}+\beta_6 (\text{\texttt{Solar.R}})^2\Big)\bigg).
\end{split}
\end{equation*}
Noting that   the response variable is now \texttt{Ozone}  and not $\log (\text{\texttt{Ozone}})$ as in  Section \ref{sub:Lin_Reg}, we modify the vector  {\tt y} accordingly:
\begin{verbatim}
R> y <- exp(y)
\end{verbatim}

The right plot in Figure \ref{figs:ozone1} shows the distribution of the observed values for the response variable \texttt{Ozone}. From this plot we observe that there is one isolated observation (i.e.~outlier), for which the observed value of   \texttt{Ozone}  is  larger than 150.

\subsubsection{GLM estimation}

We start by estimating the model with the generalized least squares (GLS) approach, using both on the full dataset and  the one obtained by removing the outlier:
\begin{verbatim}
R> glm.full <- glm(y~X,family=poisson)
R> ii <- which(y>150)
R> glm <- glm(y[-ii]~X[-ii,],family=poisson)
R> print(cbind(glm.full$coefficients,glm$coefficients))
                 [,1]         [,2]
(Intercept)  3.5168945  3.506950377
X1           2.4621470  2.332854309
X2          -0.8796456 -0.827578293
X1          -2.4753279 -2.167010600
X2           0.9498002  0.559589428
X1           4.0664346  4.289933463
X2          -0.3247376 -0.003300254
\end{verbatim}

It is well-known that the GLM estimator is sensitive to outliers and, as a result, we observe that the unique outlier present in the data has a non-negligible impact on the estimated regression coefficients of the model.

\subsubsection{MMD estimation with default setting}

We first estimate the model parameter using the  estimator $\tilde{\theta}_{\mathrm{reg}}$. Using the default setting of \texttt{mmd\_reg} this is done as follow:
\begin{verbatim}
R> mmd.tilde <- mmd_reg(y,X,model="poisson")
R> summary(mmd.tilde)
======================== Summary ========================
Model:               poisson
Estimator:           theta tilde (bdwth.x=0)
---------------------------------------------------------
  Coefficients           Estimate
---------------------------------------------------------
  (Intercept)              3.3703
  X1                       2.4924
  X2                      -1.0414
  X3                      -2.3234
  X4                       0.4451
  X5                       4.7464
  X6                      -0.1005
---------------------------------------------------------
---------------------------------------------------------
  Kernel for y: Laplace with bandwidth 18.3848
========================================================= 
\end{verbatim}

As for the linear regression example, we observe that the estimated values of the regression coefficients are similar to those obtained by GLM on the dataset without the outlier data point.

Finally, we estimate the model parameters using the estimator $\hat{\theta}_{\mathrm{reg}}$ with its default setting:
\begin{verbatim}
R> mmd.hat <- mmd_reg(y,X,model="poisson",bdwth.x="auto")
Warning message:
In mmd_reg(y,X,model= "poisson",bdwth.x ="auto") :
  Warning: The maximum number of iterations  has been reached.
\end{verbatim}
In this case, we obtain a warning message indicating that the maximum number of iterations  \texttt{maxit} allowed for the optimization algorithm has been reach. The value of \texttt{maxit}, set by default to  50\,000,  can be increased using the \texttt{control} argument of \texttt{mmd\_reg}. For instance, setting \texttt{maxit}$=10^6$ remove the warning message:
\begin{verbatim}
R> mmd.hat <- mmd_reg(y,X,model="poisson",bdwth.x="auto",
                     control=list(maxit=10^6))
R> summary(mmd.hat)
======================== Summary ========================
Model:               poisson
Estimator:           theta hat  (bdwth.x>0)
---------------------------------------------------------
  Coefficients           Estimate
---------------------------------------------------------
  (Intercept)              3.3665
  X1                        2.503
  X2                       -1.075
  X3                       -2.418
  X4                       0.2654
  X5                       4.7292
  X6                      -0.1129
---------------------------------------------------------
---------------------------------------------------------
  Kernel for y: Laplace with bandwidth 18.3848
  Kernel for x: Laplace with bandwidth 0.0215
========================================================= 
\end{verbatim}
As for the linear regression model, the value of  \texttt{bdwth.x} selected by the data driven approach implemented in \texttt{mmd\_reg} is close to zero and the estimate values of the model parameters are therefore similar to those above obtained with the estimator $\tilde{\theta}_{\mathrm{reg}}$.

\bibliographystyle{alpha}
%\setlength{\itemindent}{-\leftmargin}
%\makeatletter\renewcommand{\@biblabel}[1]{}\makeatother
\bibliography{biblio}

\end{document}